\documentclass[review]{elsarticle}

\journal{Physica A}

\usepackage{textcomp}
\usepackage{indentfirst,bm,geometry,graphicx,float,listings,xcolor,amsmath}
\usepackage{enumitem}
\begin{document}

\begin{frontmatter}

\title{Improved finite-difference and pseudospectral schemes for the Kardar-Parisi-Zhang equation with long-range temporal correlations}

\author[mymainaddress,mysecondaryaddress]{Xiongpeng Hu}

\author[mymainaddress]{Dapeng Hao\corref{correspondingauthor}}
\cortext[correspondingauthor]{Corresponding author}
\ead{hdpcumt@126.com}

\author[mymainaddress]{Hui Xia\corref{mycorrespondingauthor}}
\cortext[mycorrespondingauthor]{Corresponding author}
\ead{hxia@cumt.edu.cn}

\address[mymainaddress]{School of Materials Science and Physics, China University of Mining and Technology, Xuzhou 221116, China}
\address[mysecondaryaddress]{Software Engineering Institute, East China Normal University, Shanghai 200241, China}

\begin{abstract}
To investigate universal behavior and effects of long-range temporal correlations in kinetic roughening, we perform extensive simulations on the Kardar-Parisi-Zhang (KPZ) equation with temporally correlated noise based on pseudospectral (PS) and one of the improved finite-difference (FD) schemes.
We find that scaling properties are affected by long-range temporal correlations within the effective temporally correlated regions.
Our results are consistent with each other using these two independent numerical schemes, three characteristic roughness exponents (global roughness exponent $\alpha$, local roughness exponent  $\alpha_{loc}$, and spectral roughness exponent $\alpha_{s}$) are approximately equal within the small temporally correlated regime, and satisfy $\alpha_{loc} \approx \alpha<\alpha_{s}$ for the large temporally correlated regime, and the difference between $\alpha_{s}$ and $\alpha$ increases with increasing the temporal correlation exponent $\theta$.
Our results also show that PS and the improved FD schemes could effectively suppress numerical instabilities in the temporally correlated KPZ growth equation. Furthermore, our investigations suggest that when the effects of long-range temporal correlation are present, the continuum and discrete growth systems do not belong to the same universality class with the same temporal correlation exponent.
\end{abstract}


\end{frontmatter}


\section{Introduction}

In recent years, an enormous amount of work has been devoted to studying kinetic roughening, which has attracted much attention due to its many critical applications such as molecular beam epitaxy, bacterial growth, fluid flow in porous media, etc. \cite{Barabasi.1995}.
A few discrete models, such as the Eden, ballistic deposition (BD), and restricted solid-on-solid (RSOS) models, were proposed to describe various interface growth \cite{Meakin.1986,Eden.1961}.
Fortunately, the large-scale dynamics of such growth systems may be essentially described in terms of Langevin-type growth equations, which are also continuous approximations of various discrete models.
One of the most well-known Langevin-type growth systems is the Kardar-Parisi-Zhang (KPZ) equation \cite{Kardar.1986}, representing one important universality class of surface growth. The KPZ equation reads
\begin{equation}
\label{eq1}
\partial_{t}h(\boldsymbol{x}, t)=\nu \nabla^{2}h+\lambda(\nabla h)^{2}+\eta(\boldsymbol{x}, t),
\end{equation}
where $h(\boldsymbol{x},t)$ is the height on a $d$-dimensional substrate at position $\boldsymbol{x}$ and time $t$, $\nu$ is the effective surface tension, $\lambda$ characterizes the tilt-dependence of the growth velocity, and $\eta(\boldsymbol{x},t)$ is Gaussian noise with $\left\langle\eta(\boldsymbol{x}, t) \eta\left(\boldsymbol{x}^{\prime}, t^{\prime}\right)\right\rangle=2D\delta^{d}(\boldsymbol{x}-\boldsymbol{x}' )\delta(t-t')$. The KPZ equation satisfies Galilean invariance in any dimension, which implies there exists the hyper-scaling relation $\alpha+z=2$.

In order to conveniently perform numerical integration, Eq.~(\ref{eq1}) can be transformed by rescaling the parameters into the reduced equation with only one independent control parameter, namely coupling constant $g=\lambda \sqrt{2 D/\nu^{3}}$, which means that one can set $\nu$ and $D$ to $1$ and then replace $\lambda$ with $g$.
For a system of size $L$, the interface width $W(L,t)$ can be expressed in the dynamic scaling form\cite{Family.1985}
\begin{equation}
\label{eq2}
W(L, t)\equiv\left\langle\frac{1}{L} \sum_{i=1}^{L}\left[h\left(\boldsymbol{x}_{i}, t\right)-\bar{h}(t)\right]^{2}\right\rangle^{1/2}\sim L^{\alpha}f\left(\frac{t}{L^{z}}\right),
\end{equation}
where the overbar of $h$ denotes spatial average of the growth height, $\langle\ldots\rangle$ represents ensemble average, $\alpha$ and $z$ are the roughness and dynamic exponents, respectively.
The universal scaling function $f(u)$ has specific asymptotic properties such that $W(L,t)\sim t^{\beta}$ for $t\ll L^{z}$, and $\sim L^{\alpha}$ for $t\gg L^{z}$.
The ratio $\beta=\alpha/z$ is the growth exponent characterizing the early growth dynamics.

Since the exact solution of the KPZ equation for $d>1$ is hardly obtained due to its inherent nonlinear characteristics, the situation is even more complicated when the noise becomes correlated and modifies the universality class \cite{Barabasi.1995}.
The long-range spatiotemporally correlated noise has the form
\begin{equation}
\left\langle\eta(\boldsymbol{x}, t) \eta\left(\boldsymbol{x}^{\prime}, t^{\prime}\right)\right\rangle \sim \left|\boldsymbol{x}-\boldsymbol{x}^{\prime}\right|^{2 \rho-d}\left|t-t^{\prime}\right|^{2 \theta-1},
\end{equation}
where $\rho$ and $\theta$ are two exponents characterizing the decay of spatial and temporal correlations, respectively.
Using dynamic renormalization group (DRG) calculations, Medina et al. \cite{Medina.1989} found that, for sufficiently small $\rho$ and $\theta$, the critical exponents are the same as those for uncorrelated noise.
For $\rho$ or $\theta$ beyond a certain critical value, the correlation becomes relevant, and then the scaling exponents are nontrivial functions of $\rho$ or $\theta$.
While the effects of spatially correlated noise ($\rho \neq 0, \theta=0$) in the KPZ system have been established, such as literature \cite{Meakin.1989,HalpinHealy.1990,Zhang.1990,Amar.1991,Hentschel.1991,Peng.1991,Li.1997,Janssen.1999,Katzav.1999,MahendraK.Verma.2000,Katzav.2003,Katzav.2013,Kloss.2014,Chu.2016}, many investigations are also carried out in the KPZ equation with long-range temporal correlations ($\rho=0, \theta \neq 0$) \cite{Lam.1992,Fedorenko.2008,Strack.2015,Song.2016,Ales.2019,Squizzato.2019,Song.2021}.
Since DRG techniques have difficulty reaching the strong-coupling regimes, numerical simulations are the direct methods used to determine the universality class. It should be noted that, in dealing with the temporal correlated KPZ system, other theoretical schemes including Flory-like scaling approach (SA) \cite{ Hanfei.1993} and self-consistent expansion (SCE) \cite{Katzav.2004} are evidently inconsistent with each other as well as with DRG predictions \cite{Medina.1989}.

As a common numerical method, finite-difference (FD) scheme has been widely used to explore the continuum KPZ dynamics \cite{Moser.1991,Lam.1992,Seeelberg.1993,Song.2016,Ales.2019}.
It should be noted that, using the standard FD scheme, numerical divergence always happens in the discretized version of the KPZ equation.
In order to suppress the annoying growth instability, one can control it through replacing the square gradient of nonlinear terms by an exponentially decaying technique, as suggested by Dasgupta et al. \cite{Dasgupta.1997}.
However, the exponentially decaying function is equivalent to the nonlinear term including infinitely many higher-order nonlinearities, which may change universal behavior in comparison with the original continuum growth system \cite{Gallego.2016,Li.2021}. 
Wio et al. \cite{Wio.2010} investigated fully some discretization-related issues in the KPZ system, and discussed the relation and difference between several improved real-space discretization schemes and pseudospectral representations. Interestingly, these results show that these improved FD schemes of the KPZ equation do not obey Galilean invariance, however still remain the same universality class \cite{Wio.2010}.
Since conventional FD schemes are hampered by these discretization effects mentioned above, these unwanted numerical instabilities can be avoided (or at least delayed) in spectral schemes \cite{Gallego.2007,Giacometti.2001,Giada.2002,Priyanka.2020,Gallego.2011,Roy.2020}.
Truncating the infinite modes into finite ones, the pseudospectral (PS) approach is commonly used to deal with the nonlinear stochastic growth systems \cite{Gallego.2007,Giada.2002,Giacometti.2001,Priyanka.2020,Gallego.2011}.

Considering that the short-range correlated KPZ equation is sufficiently well studied to allow careful investigation on scaling properties, only few results are known for the long-range temporally correlated KPZ growth system, and the reliable results are still sparse. Furthermore, there are obvious differences between the existing numerical simulations and analytical approximations of the KPZ equation with long-range temporally correlated noise \cite{Medina.1989,Lam.1992,Katzav.2004, Song.2016,Ales.2019,Song.2021b}. Overall, it is necessary to further study the effects of long-range temporal correlation on scaling properties. 
In order to avoid potential risks that adopting exponentially decaying function instead of the nonlinear term may change the true universal behavior \cite{Gallego.2016,Li.2021}, the PS scheme is the preferred choice to simulate the temporally correlated KPZ growth system in this work. As an alternative method, we also adopt one of the improved FD scheme proposed by Lam and Shin \cite{LAM.1998}, hereafter referred to as the LS scheme. Using this improved FD method, the discretization of nonlinear term can effectively inhibit divergence compared with the standard central difference scheme. 
Thus, we will examine and discuss the stability and overall performance of these numerical schemes, and independently apply PS and LS numerical schemes to the temporally correlated KPZ equation clarifying universal behavior and investigating the effects of long-range temporal correlation in kinetic roughening. 

The remainder of the paper is organized as follows. First, we introduce the method to generate long-range temporally correlated noise and describe the general features of PS and LS schemes.
Then we exhibit these numerical schemes to integrate the temporally correlated KPZ equation.
After that, the characteristic quantities and scaling exponents are computed, and nontrivial scaling properties are discussed. Finally, brief conclusions and perspectives are given.

\section{Basic methods and concepts}

\subsection{Long-range temporally correlated noise}

To study such a temporally correlated growth system, we need to generate a random sequence $\left\{\eta(t)\right\}$ with a long-range power-law correlation of the form
\begin{equation}
C(\tau) \equiv\left\langle\eta(t) \eta(t+\tau)\right\rangle \sim \tau^{2 \theta-1} \quad(\tau \rightarrow \infty).
\end{equation}

For this purpose, we utilize the fast fractional Gaussian noise (FFGN) generator proposed by Mandelbrot et al. \cite{Mandelbrot.1971,Mandelbrot.1969}, adopted by Lam et al. \cite{Lam.1992} to simulate the BD model with long-range temporally correlated noise, and generalized to the temporally correlated KPZ growth systems \cite{Song.2016, Ales.2019}. It should be noted that $\left\{\eta(t)\right\}$ are independent Mandelbrot sequences for every fixed lattice site.
Using the FFGN method, the noise sequence $\left\{\eta(t)\right\}$ is defined as
\begin{equation}
\eta(t)=\sum_{n=1}^{\Lambda} W_{n} X_{t}\left(u_{n}\right),
\end{equation}
where $W_{n}$ are weight factors for different characteristic frequencies and $X_{t}$ represent Markov-Gaussian processes.
$\Lambda$ is the number of components needed, which should be increased to obtain the desired power-law exponent $\theta$ with higher precision at low frequencies.
By defining $u_{n}=a B^{-n}$, $r_{n}=e^{-u_{n}}$, $B=2$, $a=6$ and $\left\{\zeta(t)\right\}$ to be an uncorrelated sequence of uniform distribution in the interval $[0,1]$, $W_{n}$ and $X_{t}(u_{n})$ can be constructed by
\begin{equation}
\label{eq6}
\begin{aligned}
W_{n}^{2}=\frac{12\left(1-r_{n}^{2}\right)}{\Gamma(2-2 \theta)}\left(B^{\frac{1}{2}-\theta}-B^{\theta-\frac{1}{2}}\right)& u_{n}^{1-2 \theta}, \\
X_{1}\left(u_{n}\right)=\left(1-r_{n}^{2}\right)^{-0.5}\left(\zeta(1)-0.5\right) \quad t&=1, \\
X_{t}\left(u_{n}\right)=r_{n} X_{t-1}\left(u_{n}\right)+\left(\zeta(t)-0.5\right) \quad t&>1.
\end{aligned}
\end{equation}

It should be noted that another popular technique for generating long-range correlated noise is the Fourier filtering method (FFM) \cite{Chu.2016,Makse.1996}.
However, this method requires generating the complete temporal sequence of random numbers for each site, which means that the FFM takes up more memory and time than the FFGN.

\subsection{Finite-differences scheme}

Using the FD scheme, by defining the positions $x_{i}=i \Delta x(i=1, \ldots, N)$ with $N=L/\Delta x$, the spatial derivatives of the right-hand side of Eq.~(\ref{eq1}) are discretized.
For overall brevity, we define $h_{i}(t)=h\left(x_{i}, t\right)$.
Using a one-step Euler method to compute the temporal derivative, the time evolution of KPZ in $(1+1)$-dimensions reads \cite{Moser.1991}
\begin{equation}
\begin{aligned}
\label{eq7}
h_{i}(t+&\Delta t)=h_{i}(t)+\\
&\Delta t\left[\nabla^{2} h_{i}(t)+g(\nabla h_{i})^{2}(t)+\eta_{i}(t)\right],
\end{aligned}
\end{equation}
where $\Delta t$ is the time step and $\left\{\eta_{i}\left(t\right)\right\}$ is long-range correlated in the time direction generated by FFGN.

Based on LS numerical scheme \cite{LAM.1998}, the nonlinear term is discretized as
\begin{equation}
\begin{aligned}
(\nabla h_{i})^{2}(t)&=\frac{1}{3}(\Delta x)^{-2}\{\left[h_{i+1}(t)-h_{i}(t)\right]^{2} \\
&+\left[h_{i+1}(t)-h_{i}(t)\right]\left[h_{i}(t)-h_{i-1}(t)\right] \\
&+\left[h_{i}(t)-h_{i-1}(t)\right]^{2}\}.
\end{aligned}
\end{equation}
Notably, it has been argued that this LS discretization can recover the results of the continuum growth equations while discrepancies arise in the use of conventional discretization \cite{LAM.1998}.

\subsection{Pseudospectral scheme}

PS scheme is another numerical scheme we adopted in this work, in which the spatial gradients are computed using the fast Fourier transform (FFT).
Assuming that the height field $h(\boldsymbol{x},t)$ satisfies periodic boundary conditions in the interval $[0,L]$, it can be described by Fourier modes $\hat{h}(\boldsymbol{q}, t)$
\begin{equation}
\label{eq11}
h(\boldsymbol{x}, t)=\frac{1}{L} \sum_{k=-\frac{N}{2}}^{\frac{N}{2}-1} \hat{h}\left(q_{k}, t\right) e^{i q_{k} x}=\mathcal{F}^{-1}[\hat{h}(\boldsymbol{q}, t)],
\end{equation}
where $q_{k}=\frac{2 \pi}{L} k$ represents the wave number.
Applying FFT to the temporally correlated KPZ equation, we obtain a set of ordinary differential equations (ODEs)
\begin{equation}
\frac{d \hat{h}_{k}(t)}{d t}=\hat{\mathcal{L}}_{k}(t)+g\widehat{\mathcal{N}}_{k}(t)+\hat{\eta}_{k}(t).
\end{equation}
The temporal discretization for the above complex equations is performed by a one-step Euler method.
It is easy to verify that the Fourier modes $\hat{\eta}_{k}(t)$ is derived from the Fourier transform of $\eta(\boldsymbol{x}, t)$ with the correlations
\begin{equation}
\left\langle\hat{\eta}_{k}(t) \hat{\eta}_{k^{\prime}}\left(t^{\prime}\right)\right\rangle \sim \delta_{k,-k^{\prime}}\left|t-t^{\prime}\right|^{2 \theta-1}.
\end{equation}

The linear term $\hat{\mathcal{L}}_{k}(t)$ is the Fourier transform of $\nabla^{2}h$, which is given by
\begin{equation}
\hat{\mathcal{L}}_{k}(t)=-q_{k}^{2} \hat{h}_{k}(t).
\end{equation}
Using PS approach, aliasing appears when dealing with the nonlinear term, and the wave number in convolution exceeds the frequency range.
The nonlinear term $\widehat{\mathcal{N}}_{k}(t)$ reads
\begin{equation}
\label{eq15}
\widehat{\mathcal{N}}_{k}(t)=-\left [ \sum_{k_{1}+k_{2}=k}^{}+ \sum_{k_{1}+k_{2}=k\pm N}^{} \right ]q_{k_{1}}q_{k_{2}}\hat{h}_{k_{1}}(t)\hat{h}_{k_{2}}(t).
\end{equation}
The first sum in Eq.~(\ref{eq15}) is the right convolution sum whereas the second sum is called aliasing error \cite{Gallego.2007,Giada.2002,Gallego.2011}. Here, we use zero-padding approach to eliminate the aliasing error. That is, to make the $N$ Fourier modes in the above centered convolution free of aliasing, we extend the vector on both sides by zeros to a larger wave vector of $k$. The minimum and most efficient choice for $k$ turns out to be ${3N}/{2}$, which is discussed in detail in Ref.~\cite{Giada.2002}.

\subsection{Information on numerical simulations}

For simulating the temporally correlated KPZ equation, we find that numerical divergences are very evident using the standard FD scheme. Therefore, it is necessary to adopt appropriate numerical methods. PS could control successfully numerical divergences in dealing with various local growth systems, including KPZ with Gaussian white noise \cite{Gallego.2007,Giada.2002,Gallego.2011}, adopting PS scheme becomes our first choice for simulating the temporally correlated KPZ equation.
For complete comparison of the stability and reliability of numerical methods, the LS scheme is also adapted to perform numerical investigations on this kind of generalized KPZ system. 

In our simulations, the time evolution of the interface is started from an initially flat profile $h(x,0)=0$ with periodic boundary conditions. We use spatial and temporal discretization of $\Delta x=1$ and $\Delta t=0.05$. To obtain the expected correlated noise from Eq.~(\ref{eq6}) with different temporal correlation exponent $\theta$, we set the value of $\Lambda$ to $50$ for high precision. Thus, we only need to adjust the values of nonlinear coefficient $g$ for a given temporal correlation exponent. Specifically, $g=1.41$ for PS, and $g=1.69$ for LS are chosen to ensure the simulated system into the true nonlinear KPZ scaling regime. 

When performing independently the numerical simulations based on PS and  LS schemes, we could reproduce the spontaneous formation of faceted patterns, which was first reported in Ref. \cite{Ales.2019}, and these faceted profiles become apparent in the long time limit as the exponent $\theta$ is increased. We also observe that there exists evident change of the height profiles for $\theta$ beyond a certain critical threshold $\theta_{th}$ from self-affine profiles to faceted patterns \cite{Ales.2019,Song.2021b}.

\section{Numerical results and discussions}

To quantitatively describe the scaling properties, we obtain the growth exponent $\beta$ and roughness exponent $\alpha$ with different $\theta$ by computing the time evolution of interface width $W(L,t)$.
Considering the fluctuations of $W(L,t)$ with system size and growth time, we use the large system sizes and noise realizations.

We calculate an effective growth exponent $\beta(t)$ defined as \cite{Kelling2011}

\begin{equation}
\label{eqbeta}
\beta(t)=\frac{ln[W(L,t)/W(L,t')]}{ln(t/t')},
\end{equation}
using $t/t'=2$. We find that, for a given $\theta$, similar scaling curves could be plotted with different system sizes from $L=2048$ to $16384$, which means that finite-size effects are relatively weak within early growth times. Then, we obtain the estimated values of growth exponent $\beta$ by averaging the early obvious platform stages of these effective $\beta(t)$ curves from the largest system size  $L=16384$ to reduce fluctuations \cite{Kelling2011}. The more detailed presentation for the estimated values of  $\beta$ is shown in Appendix, and the comparison of local slopes versus $1/t^{1/2}$ and $t$ for different system sizes are shown in Figs.A1 and A2, respectively. 
However, in the saturated growth regime $t\gg L^{z}$, previous studies on simulating discrete and continuum growth systems show that finite-size effects should be considered to obtain the effective estimated values of $\alpha$ \cite{AaraoReis.2001, AaraoReis.2006a, Miranda.2008, Pagnani.2015}.
Finite-size estimates of the roughness exponents are given by
\begin{equation}
\label{eqalphaL}
\alpha(L)=\frac{ln[W(L,t\to\infty)/W(L/2,t\to\infty)]}{ln2}.
\end{equation}
The effective exponent $\alpha$ is obtained form scaling form\cite{AaraoReis.2001}:
\begin{equation}
\label{eqalpha}
\alpha(L)\approx \alpha+AL^{-\Delta_1 },
\end{equation}
where $\Delta_1$ is a correction-to-scaling exponent, and $A$ is a constant.

\begin{figure}
\centering
\includegraphics[scale=0.65]{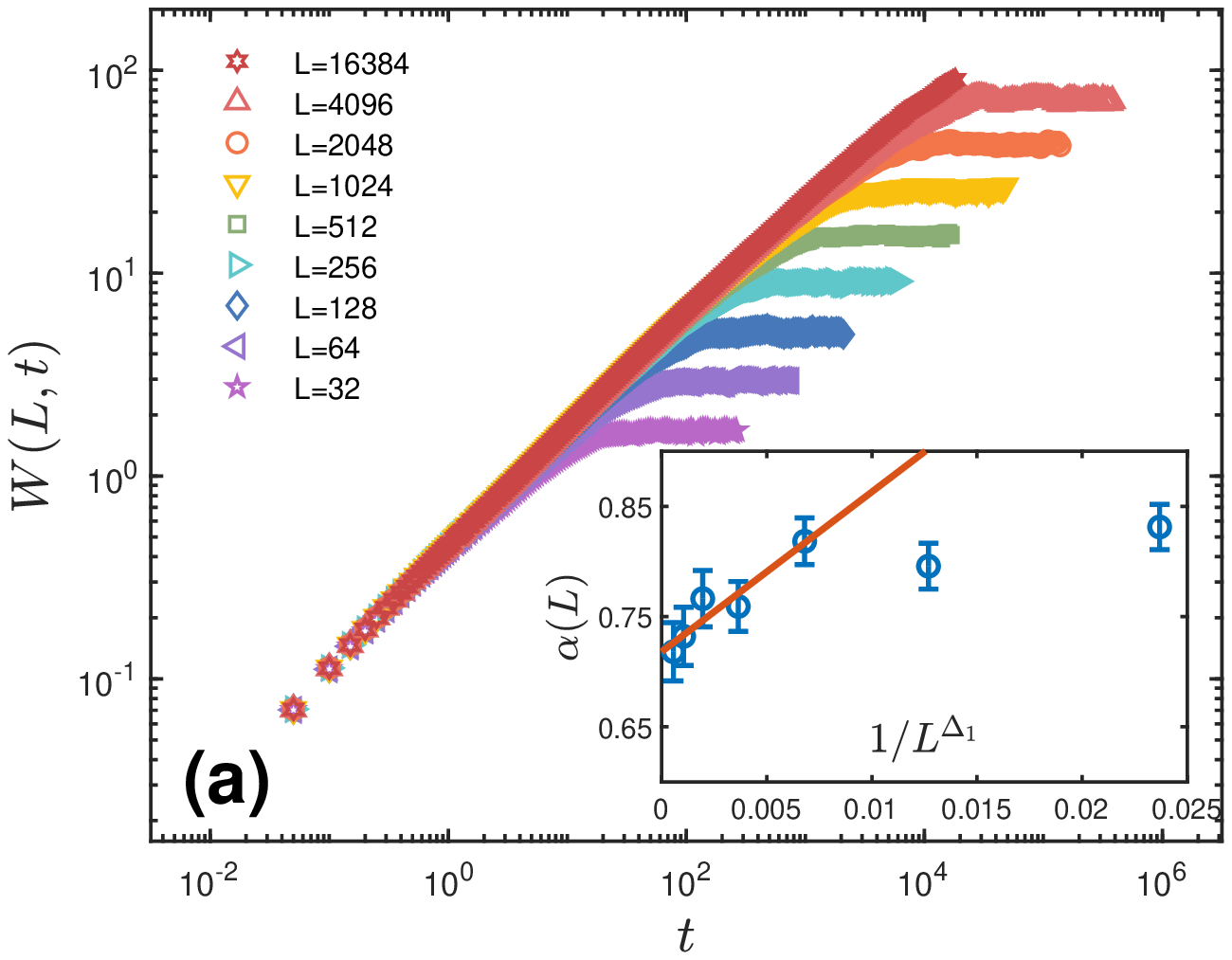}
\includegraphics[scale=0.65]{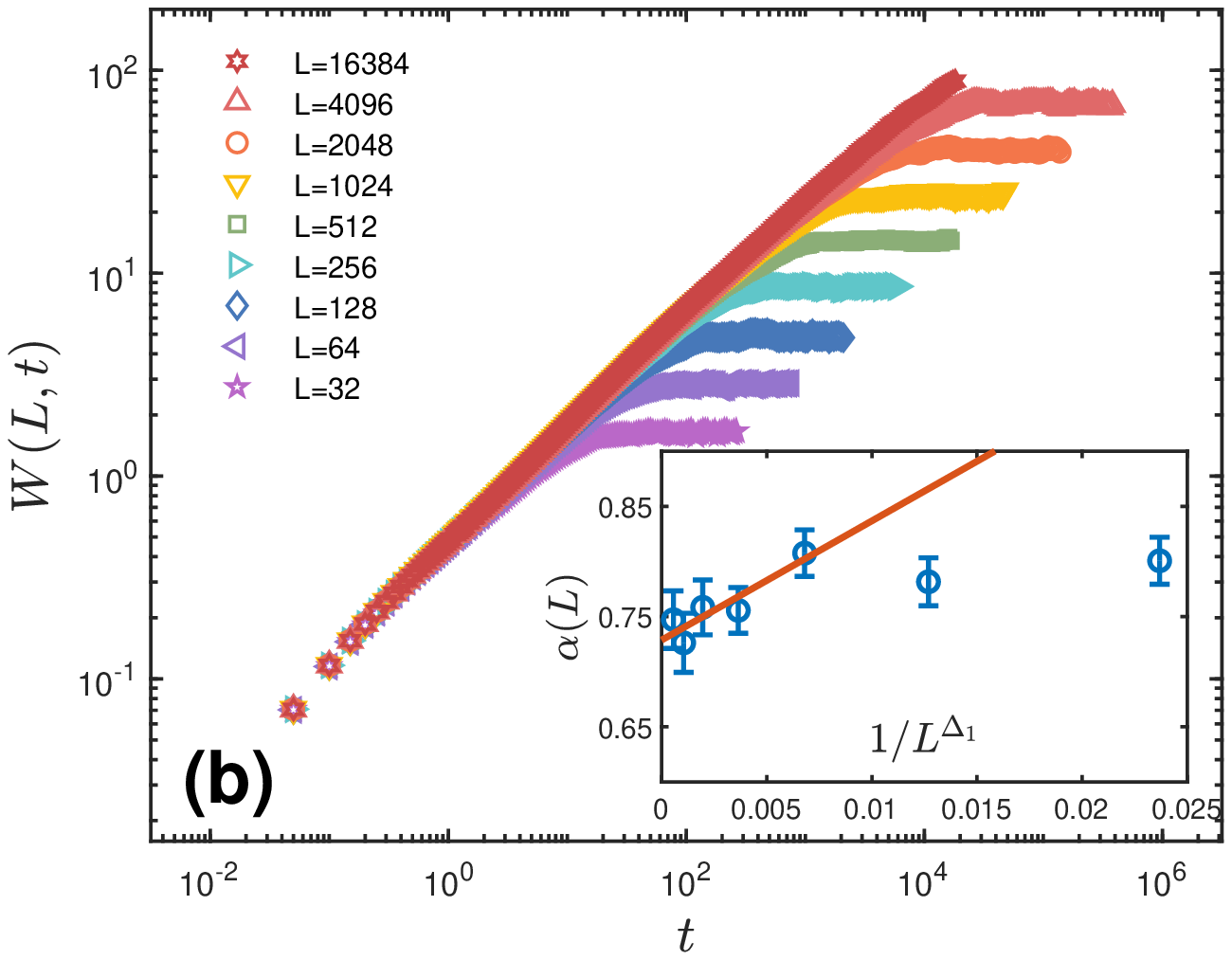}
\caption{\label{fig1}The plot of $W(L,t)$ at different growth regimes for $\theta=0.30$ using these two schemes: (a) PS and (b) LS. The dashed lines are the fitting results of $\beta$. The insets exhibit effective estimated values of global roughness exponent $\alpha$ as a function of inverse system size $1/L^{\Delta_1}$ with $\Delta_1=0.9$. Data were averaged over $750$ independent noise realizations.}
\end{figure}

In our simulations, we calculate the scaling exponents using both PS and LS schemes with $\theta$ varying in the interval $[0,1/2)$.
In Fig.~\ref{fig1}, we plot the independent results using these two numerical schemes. Here $\theta=0.30$ is chosen as a typical representative.
In the early time regime $t\ll L^{z}$, we obtain the growth exponents 
$\beta=0.583\pm 0.008$ 
for the PS scheme, and 
$\beta=0.580\pm 0.010$ 
for LS case.
In the saturated growth regime $t\gg L^{z}$, using finite-size corrections \cite{AaraoReis.2001, AaraoReis.2006a}, we obtain the global roughness exponents 
$\alpha=0.73\pm 0.03$
for PS and 
$\alpha=0.73\pm 0.03$
for LS scheme, as shown in Fig.~\ref{fig1}. Thus, we find that both PS and LS schemes give almost identical results, which are in good agreement with DRG predictions ($\alpha  \approx 0.727$ for $\theta=0.30$) \cite{Medina.1989}.
Considering that the scaling exponents $\alpha$, $\beta$, and $z$ are not independent, and there exists a simple way to collapse the data onto a single curve.
We could rescale the interface width $W$ by $W/L^{\alpha}$ vertically and rescale growth time $t$ by $t/L^{z}$ horizontally.
Our results could also show good collapses as similar as Ref.~\cite{Song.2021b}, which means that we obtain effectively the values of $\alpha$ and $z$.

\begin{figure}
\centering
\includegraphics[scale=0.65]{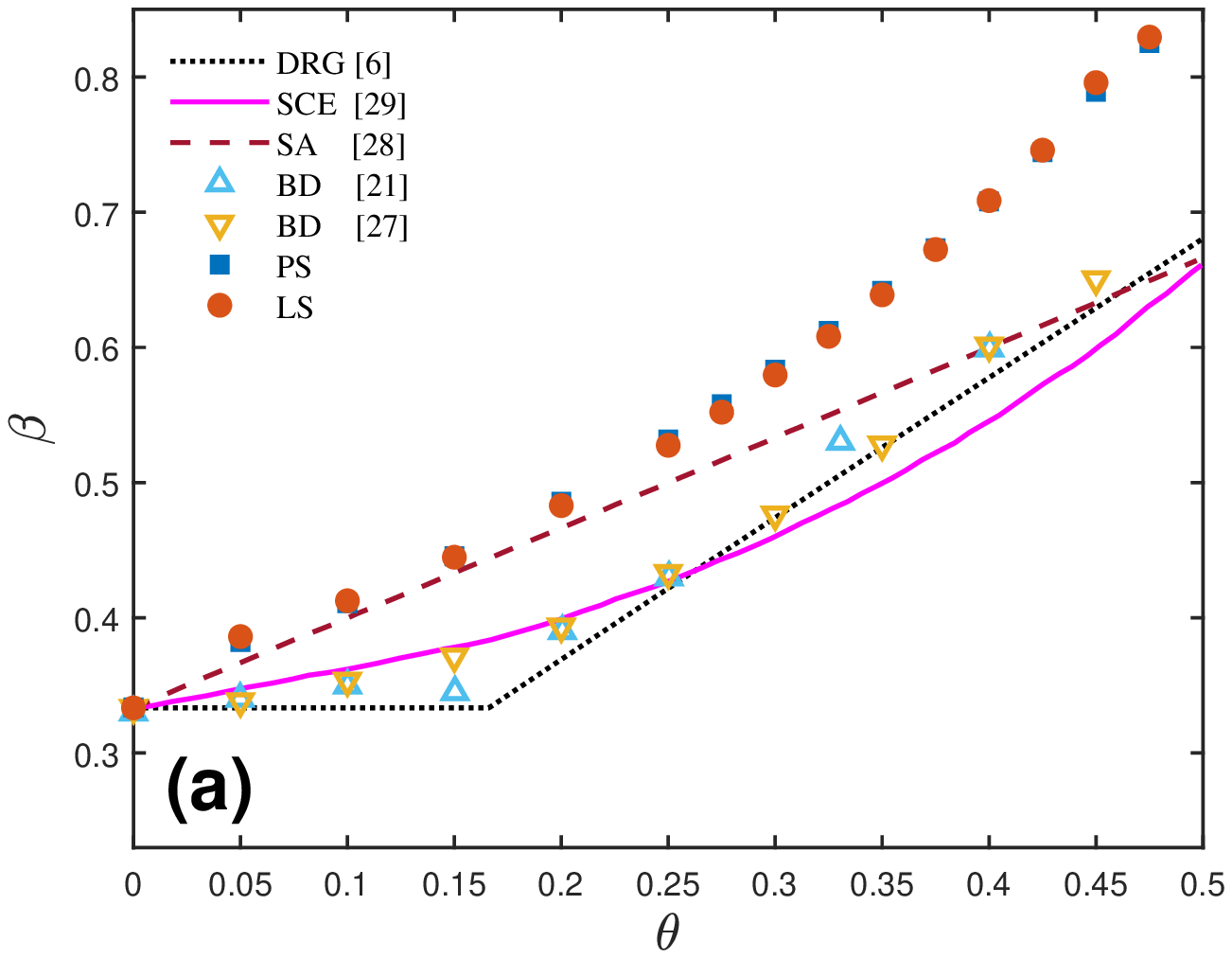}
\includegraphics[scale=0.65]{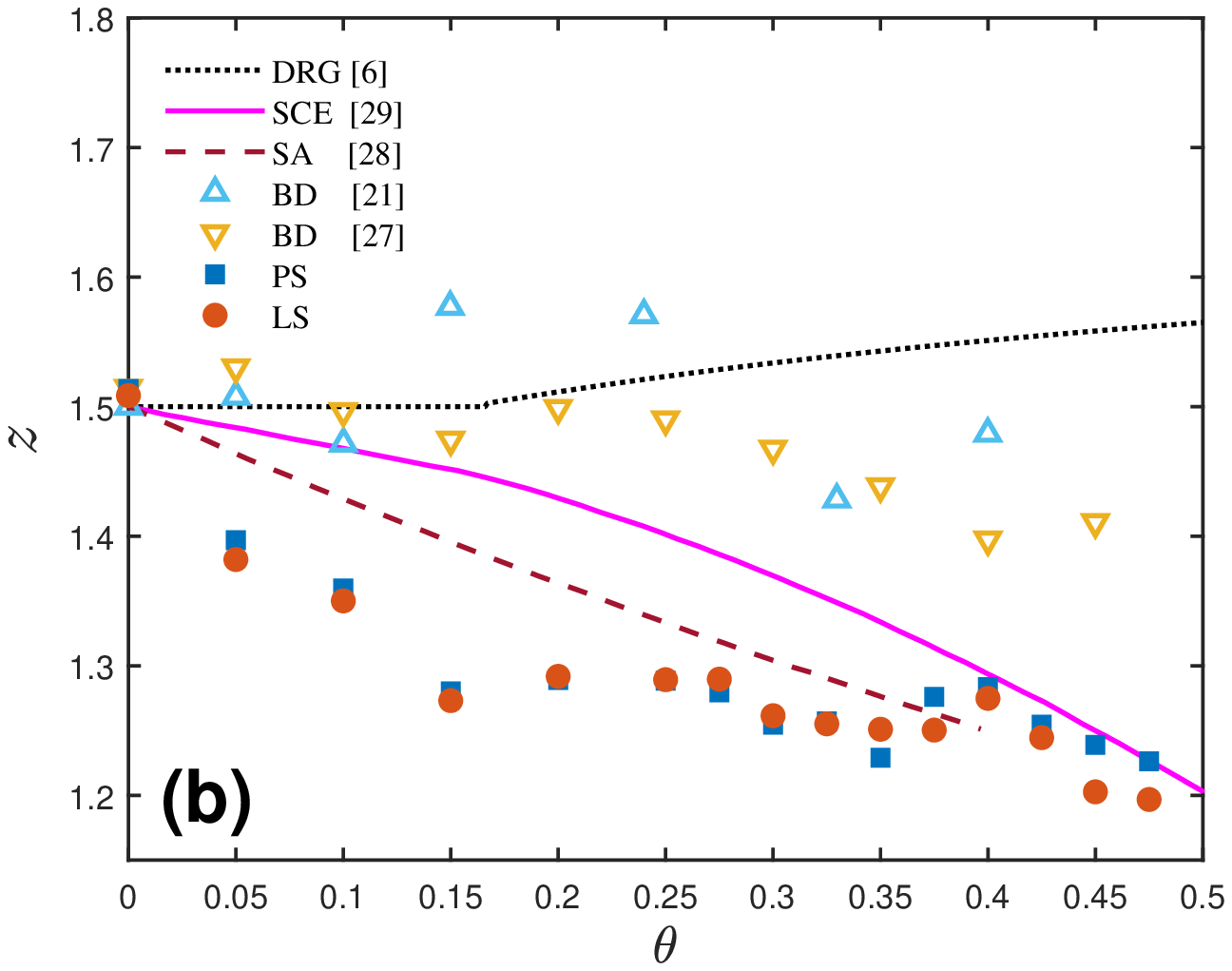}
\caption{\label{fig2}(a) The growth exponent $\beta$, and (b) the dynamic exponent $z$ as a function of the exponent $\theta$ for decay of temporal correlations. Note that $z$ was obtained using the scaling relation $z=\alpha /\beta$.}
\end{figure}

In Fig.~\ref{fig2}, our results show that the growth exponent varies distinctly with $\theta$.
The scaling behavior has a certain dependence on the temporal correlation exponent in the early growth regime.
We find that the results calculated by the PS scheme are very close to those by the LS method, demonstrating that our numerical results based on these two schemes are in good agreement with each other in simulating the KPZ equation with long-range temporal correlations. When $\theta$ approaches $0$, these results can be reduced to the normal KPZ equation based on PS and LS schemes. Compared with the previous theoretical predictions \cite{Medina.1989,Katzav.2004,Hanfei.1993} and numerical simulations \cite{Lam.1992,Song.2021}, the growth exponents in these investigations are different. Remarkably, $\beta$ obtained here is larger than the results from simulating the BD model in the presence of temporally correlated noise.
Thus, our results imply that when the temporal correlation exponent is above a certain threshold, the effects of long-range temporal correlation are not ignored. In this case, the continuum KPZ and discrete BD growth systems do not belong to the same universality class in the presence of the same temporal correlation exponent. Interestingly, Peng et al. \cite{Peng.1991} suggested that BD driven by long-range spatially correlated noise does not belong to the spatial correlated KPZ universality class.
The difference could be due to the different correlated noise generation methods between KPZ and BD systems. One needs to binarize the correlated noise into 0 or 1 in the modified BD system, while we need not introduce this binarization in simulating temporally or spatially correlated KPZ equations.

\begin{figure}
\centering
\includegraphics[scale=0.65]{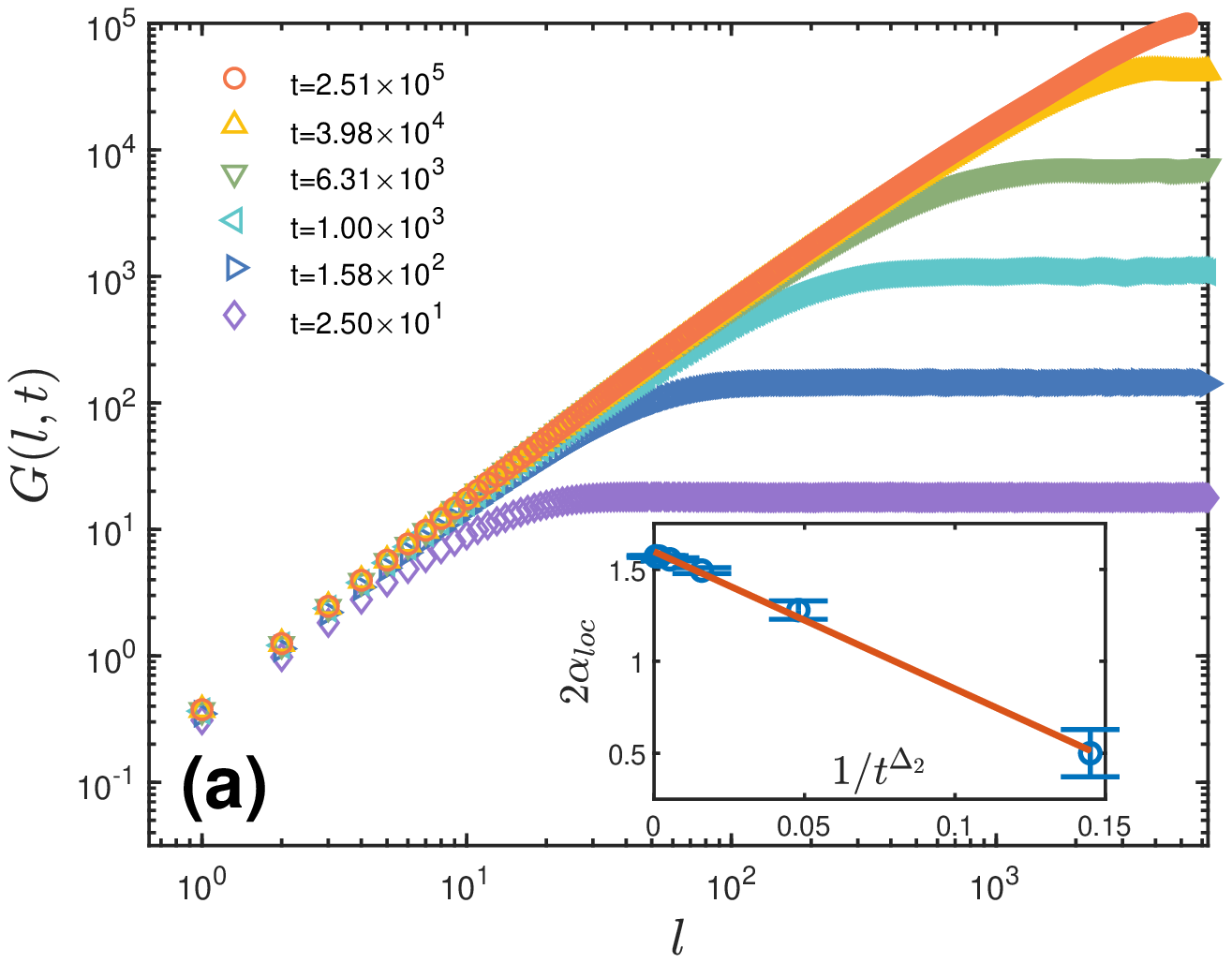}
\includegraphics[scale=0.65]{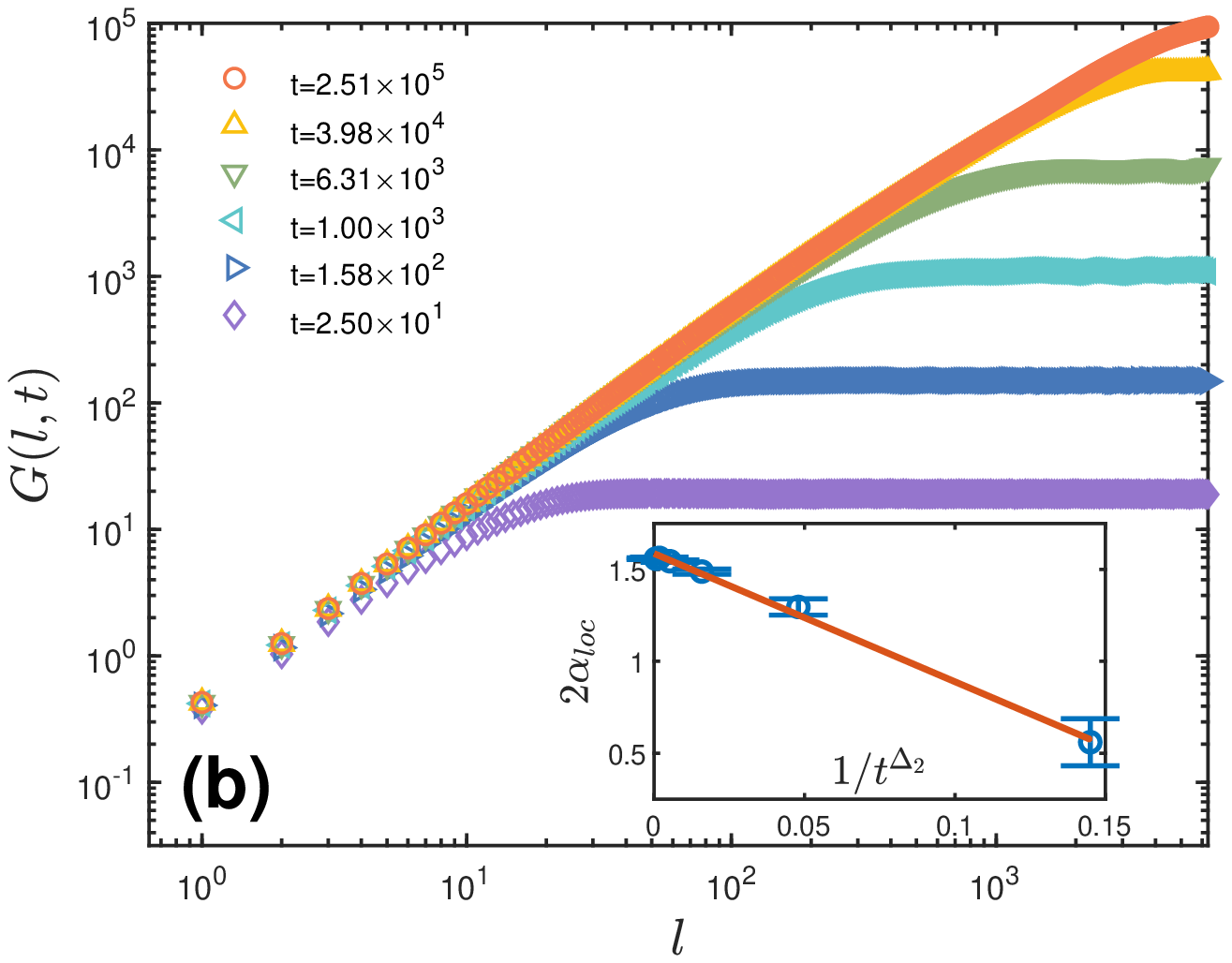}
\caption{\label{fig3}The plot of height difference correlation function $G(l,t)$ versus $l$ for $\theta=0.30$ based on (a) PS and (b) LS schemes. Here $L=16384$ and data were averaged over $750$ independent noise realizations. The insets exhibit the effective estimated values of local roughness exponent $\alpha_{loc}$ based on finite-time corrections with $\Delta_2=0.6.$}
\end{figure}

Kinetically rough surfaces generated by Eq.~(\ref{eq1}) generally possess scale-invariant properties.
However, the existence of power-law scaling of the correlation functions does not determine a unique dynamic scaling form, which leads to the different anomalous forms of dynamic scaling \cite{Ramasco.2000}.
One more independent exponent may be needed to assess the universality class of the particular stochastic growth system under study.
Here, we recall some characteristic quantities in which the scaling behavior becomes manifest as a power law and define the relevant critical exponents.
The equal-time height difference correlation function $G(l,t)$ is the correlated quantity between height fluctuations between any two points $x,x+l$, which reads \cite{Lopez.1997}
\begin{equation}
G(l,t) = \left \langle \left ( h(x+l,t)-h(x,t) \right )^{2}  \right \rangle.
\end{equation}
The brackets $\left \langle \dots \right \rangle$ denote average over noise, and the local roughness exponent $\alpha_{loc}$ is determined from the relation $G(l,t)\sim l^{2\alpha_{loc}}\quad(l\ll L)$.
A normal self-affine interface satisfies $\alpha=\alpha_{loc}$, and $\alpha \ne \alpha_{loc}$ indicates anomalous scaling in surface growth.
As shown in Fig.~\ref{fig3}, we exhibit the plots of $G(l,t)$ versus $l$ with different growth times when $\theta=0.30$. To suppress possible finite-time effects, we perform finite-time corrections to obtain the asymptotic value of $\alpha_{loc}$ when the growth time tends to infinity using $\alpha_{loc}(t)\approx \alpha_{loc}+Bt^{-\Delta_2}$.
We obtain 
$\alpha_{loc}=0.788+0.005$
for the PS scheme, and 
$\alpha_{loc}=0.783+0.005$
for LS scheme. It is found that,  for the whole $\theta$ region,  the values of $\alpha_{loc}$ and $\alpha $ are always approximately equal.
And detailed comparisons between $\alpha_{loc}$ and $\alpha$ versus $\theta$ will be provided in the following. Notedly, recent numerical investigations revealed that, through direct simulating the temporally correlated KPZ equation with the exponentially decaying function, there is no obvious power-law relationship from equal-time height difference correlation function, which implies $\alpha_{loc}$ exhibiting nonuniversal scaling within local window sizes \cite{Song.2021b}. Therefore, the exponentially decaying function could effectively suppress numerical divergence in simulating the local KPZ equation; it should be cautious about adopting this function instead of the nonlinear term in the KPZ growth system driven by long-range temporally correlated noise.

Apart from the height difference correlation function, one of the most general descriptions of the scaling properties in kinetic roughening is the structure factor, which reads \cite{Ramasco.2000}
\begin{equation}
S(k,t)=\left \langle \hat{h}_{k}(t)\hat{h}_{-k}(t) \right \rangle,
\end{equation}
where $\hat{h}_{k}(t)$ is the discrete Fourier coefficients obtained by Eq.~(\ref{eq11}).
For $(1+1)$-dimensions, the structure factor satisfies the scaling form
\begin{equation}
S(k,t)=k^{-(2\alpha+1)}s(kt^{1/z}),
\end{equation}
where the most general scaling function $s(u)$, consistent with scale-invariant dynamics, is given by \cite{Ramasco.2000}
\begin{equation}
s(u)\sim\begin{cases}
u^{2(\alpha-\alpha_{s})}\quad &u\gg 1, \\
u^{2\alpha+1}\quad &u\ll 1.
\end{cases}
\end{equation}
Here, $\alpha_{s}$ is the spectral roughness exponent. It is important to remark that $S(k,t)$ allows one to obtain the distinctively characteristic spectral roughness exponent $\alpha_{s}$ typical of faceted growing surface, which leaves no trace in the usual height difference correlation functions \cite{Ales.2019}.
Standard scaling corresponds to $\alpha_{loc}=\alpha_{s}=\alpha<1$.
However, other situations may be described within the generic scaling framework, including super-roughening ($\alpha_{loc}=1,\alpha_{s}=\alpha>1$), intrinsic anomalous scaling ($\alpha_{loc}=\alpha_{s}<1,\alpha_{s}\ne \alpha$) and faceted scaling ($\alpha_{s}>\alpha_{loc}=1, \alpha_{s}\ne \alpha$) \cite{Ramasco.2000}.
To determine the universal critical properties, we utilize the power-law of structure factor versus wave number. 

\begin{figure}
\centering
\includegraphics[scale=0.65]{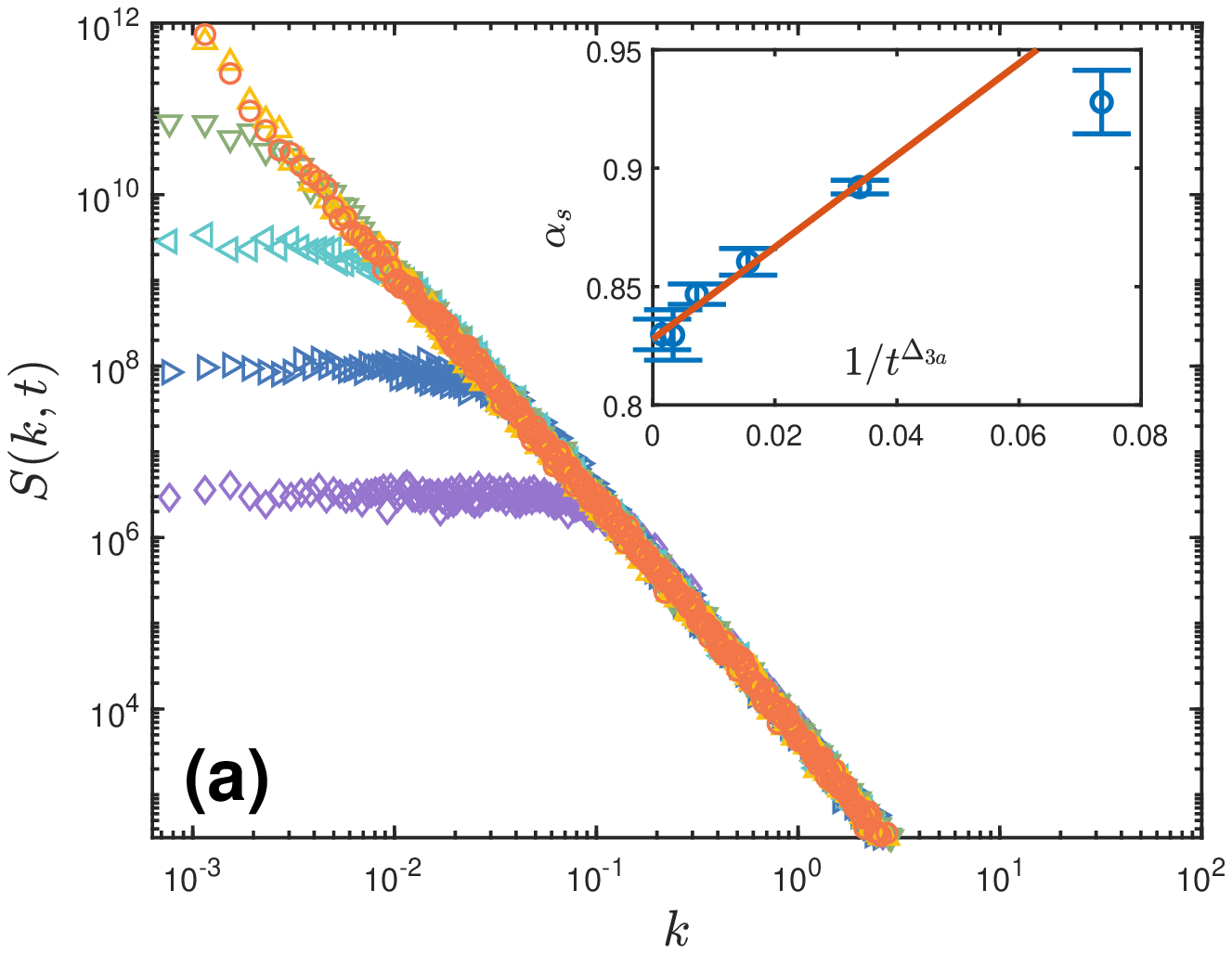}
\includegraphics[scale=0.65]{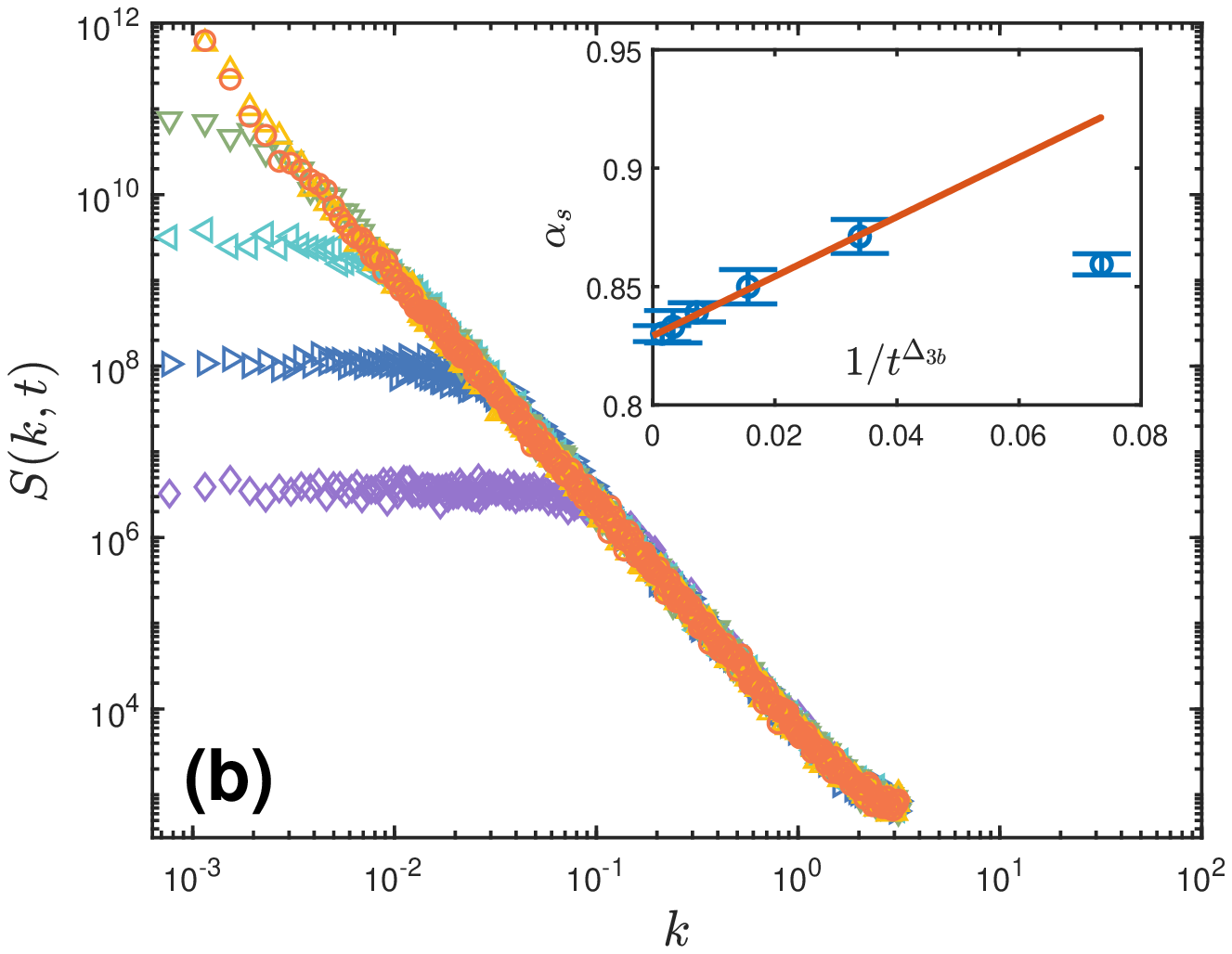}
\caption{\label{fig4}The plot of structure factor $S(k,t)$ versus $k$ for $\theta=0.30$ using (a) PS and (b) LS schemes. The insets exhibit the effective estimated values of spectral roughness exponent $\alpha_{s}$ based on finite-time corrections  with $\Delta_{3a}=0.41$ and $\Delta_{3b}=0.42$. Growth times and the independent noise realizations are the same as those in Fig.~\ref{fig3}.}
\end{figure}

\begin{figure}
\centering
\includegraphics[scale=0.65]{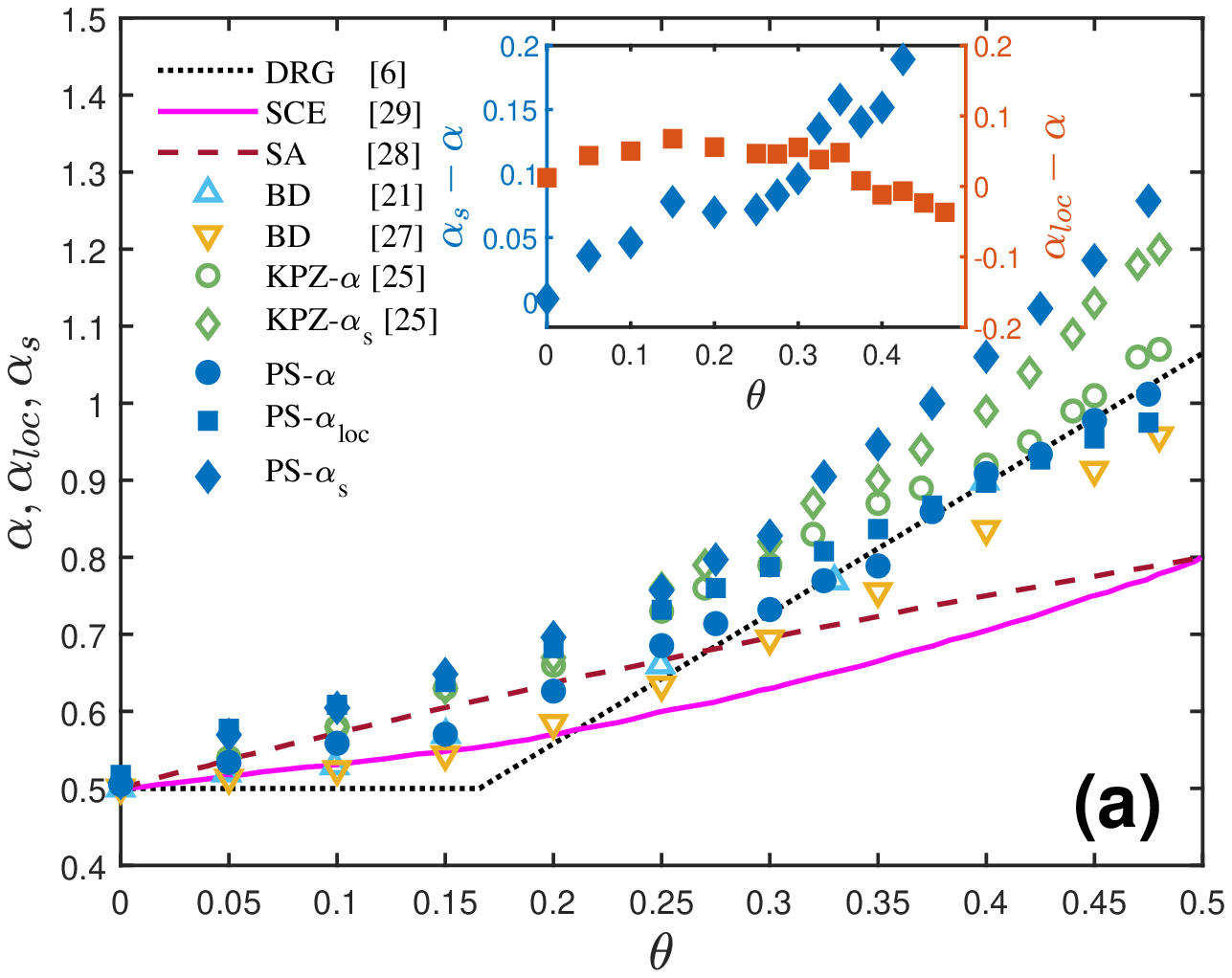}
\includegraphics[scale=0.65]{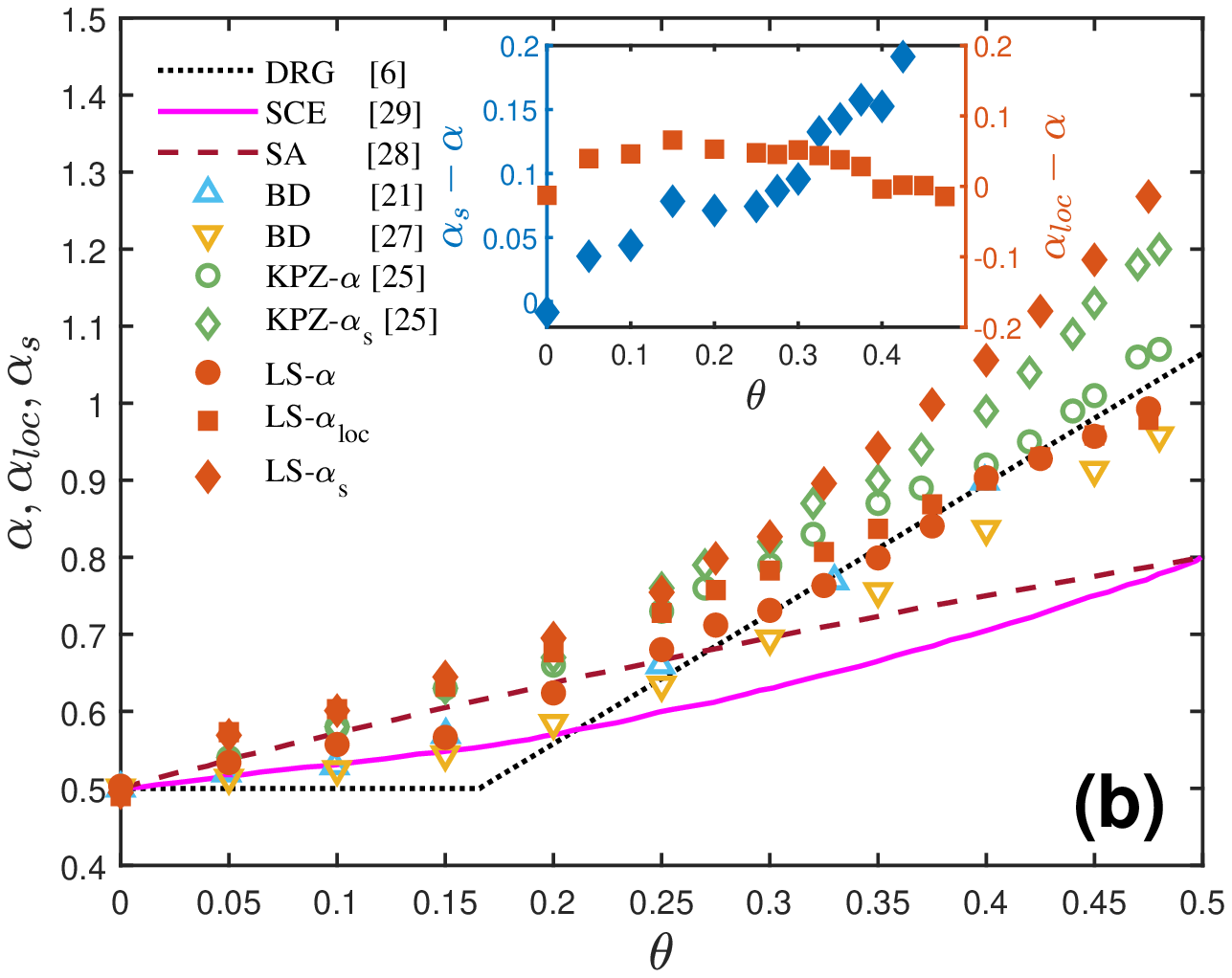}
\caption{\label{fig5}The scaling exponents $\alpha,\alpha_{loc},\alpha_{s}$ versus $\theta$ with (a) PS and (b) LS schemes. The existing predictions and numerical results of $\alpha$ are also provided for comparison quantitatively. The insets show the $\alpha_{s}-\alpha$ (diamonds), and $\alpha_{loc}-\alpha$ (squares) versus $\theta$ correspondingly.}
\end{figure}

Account for possible finite-time effects, we also adopt the extrapolation of spectral roughness exponents with different growth times based on using $\alpha_{s}(t)\approx \alpha_{s}+Ct^{-\Delta_3}$. Figure~\ref{fig4} exhibits the plot of structure factor $S(k,t)$  when $\theta=0.30$.
We obtain 
$\alpha_{s}=0.828+0.006$
for PS scheme and 
$\alpha_{s}=0.827+0.006$
for LS method based on asymptotic estimates with the spectral roughness exponent, as shown in the insets of Fig.~\ref{fig4}.
We find that the effective estimated values of $\alpha_s$ here are larger than the actual calculated values of $\alpha$ and $\alpha_{loc}$ within the large $\theta$ region, as shown in Fig.~\ref{fig5}.
Through the full comparison of the scaling exponents within the whole $\theta$ region, we find that these estimated values of $\alpha$ and $\alpha_{s}$ could be consistent with the previous numerical results of simulating the KPZ driven by long-range temporally correlated noise \cite{Ales.2019,Song.2021b}. The evident difference between $\alpha$ and $\alpha_{s}$ implies anomalous scaling appearing in the temporally correlated KPZ system.
For the global roughness exponent, we also find that  $\alpha$ versus $\theta$ displays roughly the linear relationship for the small $\theta$ region, and our results are in well agreement with DRG predictions for the large $\theta$ region.
Another interesting finding is that the slope of $S(k,t)$ becomes smaller using the PS scheme, but a little larger for adopting LS when $kt^{1/z}\gg 1$.
The difference between  $\alpha_{s}$ and $\alpha$ becomes smaller as $\theta \to 0$, meaning that the temporally correlated KPZ equation approaches the normal KPZ equation.

\section{Conclusions}

We performed numerical investigations of the KPZ equation with long-range temporally correlated noise based on the PS approach and LS discretization, one of the improved FD schemes.
Long-range temporal correlations evidently affect the scaling properties of the temporally correlated KPZ system.
Spontaneous formation of faceted patterns \cite{Ales.2019,Song.2021b} becomes apparent as the temporal correlation exponent $\theta$ increases.
As for scaling exponents $\alpha$, $\beta$, and $z$, we find that the results of PS and LS schemes are consistent with each other, demonstrating that these two numerical schemes we adopted are reliable in simulating KPZ equations with long-range temporal correlations.
When $\theta$ approaches $0$, both PS and LS schemes could obtain the results of the normal KPZ equation with Gaussian white noise.
Our results show that the global roughness exponent $\alpha$ versus $\theta$ is approximately linear for the small $\theta$ region, and agrees with DRG for the large $\theta$ region, which are in agreement with the previous numerical results \cite{Ales.2019, Song.2021b}.
 This investigation's growth exponent $\beta$ is higher than previous numerical studies based on BD in the presence of long-range temporal correlations \cite{Song.2021}. The dynamic exponent $z$ roughly matches the SCE predictions, which also differs from the previous results of direct simulating BD driven by temporally correlated noise. Thus, our results show that, even in the presence of the same temporally correlated noise, the scaling exponents $\beta$ and $z$ from directly simulating KPZ are slightly different from those obtained from the temporally correlated BD model. As a qualitative explanation, the difference between the discrete and continuum growth systems is due to two different ways of generating correlated noise sequences. One has to binarize the correlated noise into 0 or 1 in the temporally correlated BD system.
By comparing these three characteristic roughness exponents $\alpha_{s}$,  $\alpha_{loc}$ and $\alpha$, we find that they are approximately equal in the small $\theta$ regions, and $\alpha_{loc} \approx \alpha < \alpha_{s}$ in the large $\theta$ regions. More precisely, the difference between $\alpha_{s}$ and $\alpha$ becomes evidently larger as $\theta$ increases. And the estimated values of $\alpha_{loc}$ and $\alpha$ are approximately equal in the whole $\theta$ regime. Thus, the scaling relationships mentioned above could fit roughly into the faceted scaling of generic scaling theory \cite{Ramasco.2000}.

It should be noted that numerical divergences always exist in simulating the temporally correlated KPZ equation using different numerical schemes. The leading cause of divergence is the discretization of the nonlinear term. Through the complete comparison, we find that adopting PS and LS schemes all could maintain evidently more long growth times than those using the standard FD methods. Fortunately, within the reasonable system sizes and growth times, the interface width could steadily reach the saturated regime using these two numerical schemes before numerical instability appears. Thus we could successfully obtain various scaling exponents.
Interestingly, based on the PS approach, this kind of numerical instability could be avoided entirely in simulating the local growth equations, reported in the previous literature \cite{Gallego.2007,Giacometti.2001,Giada.2002,Priyanka.2020,Gallego.2011}. Therefore, the range of validity and effectiveness of PS scheme remains elusive, and needs to investigate utilizing various nonlinear stochastic equations, especially those including nonlocal interactions or long-range correlations. Thus, this work also motivates investigating alternative numerical methods for dealing with nonlinear stochastic systems with long-range spatial and temporal correlations.
Furthermore, height distribution and the asymptotic growth speed $v_{\infty}$ have been introduced to provide a more detailed characterization of kinetic roughening processes \cite{Takeuchi.2013, Takeuchi.2018}. How to affect the height distribution and the asymptotic growth speed with adjusting temporal correlation exponent should be investigated in the continuum and discrete growth systems with long-range temporal correlation. Undoubtedly, further study of this issue would be of interest.

\section* {Acknowledgements}
This work was supported by the Fundamental Research Funds for the Central Universities under Grant No.2020ZDPYMS31.



\bibliography{mybibfile}

\appendix 

\section*{Appendix: Detailed presentation for estimating the values of growth exponent when finite-size effects are not obvious within the early growth regions}

\begin{figure}
\centering
\includegraphics[scale=0.65]{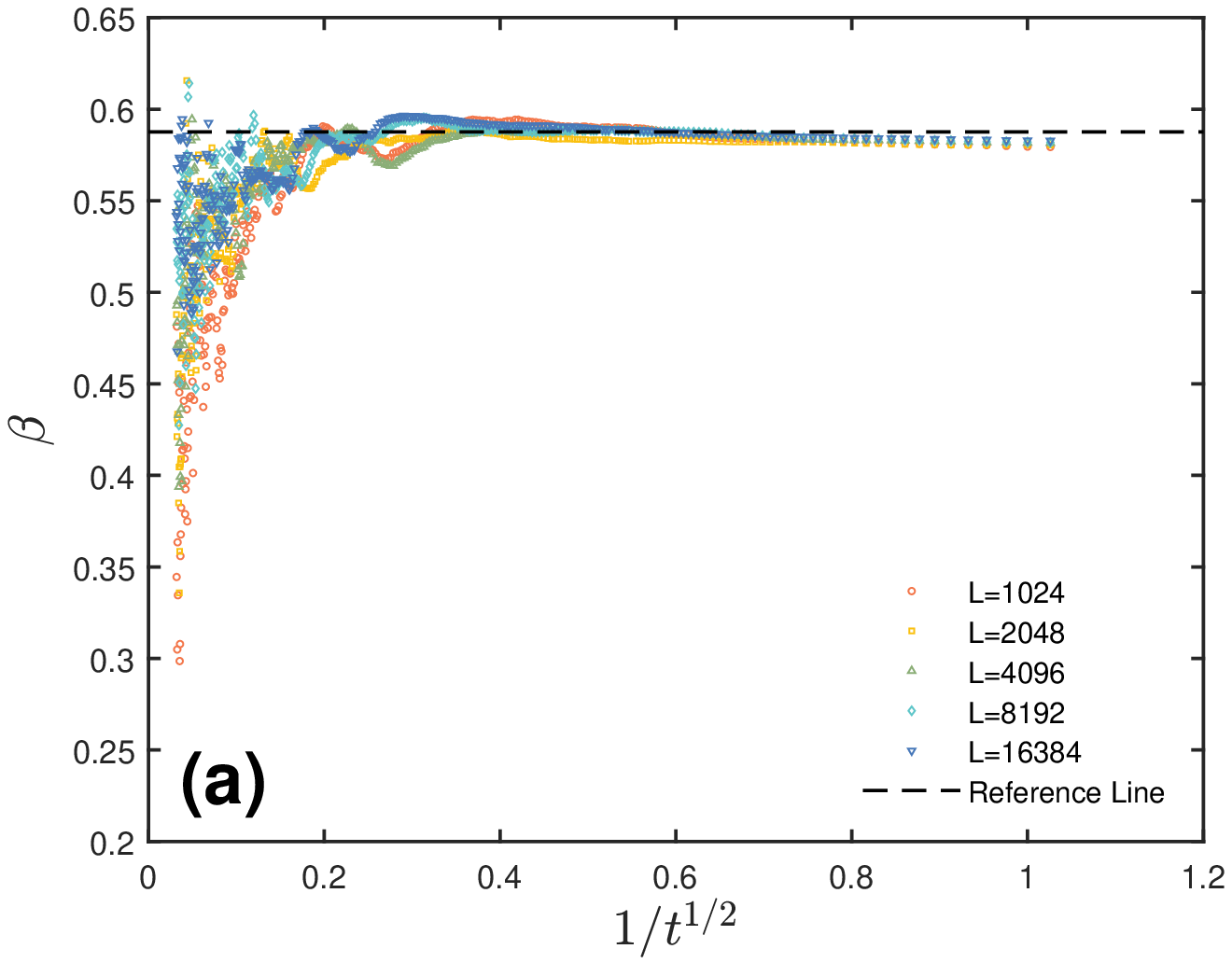}
\includegraphics[scale=0.65]{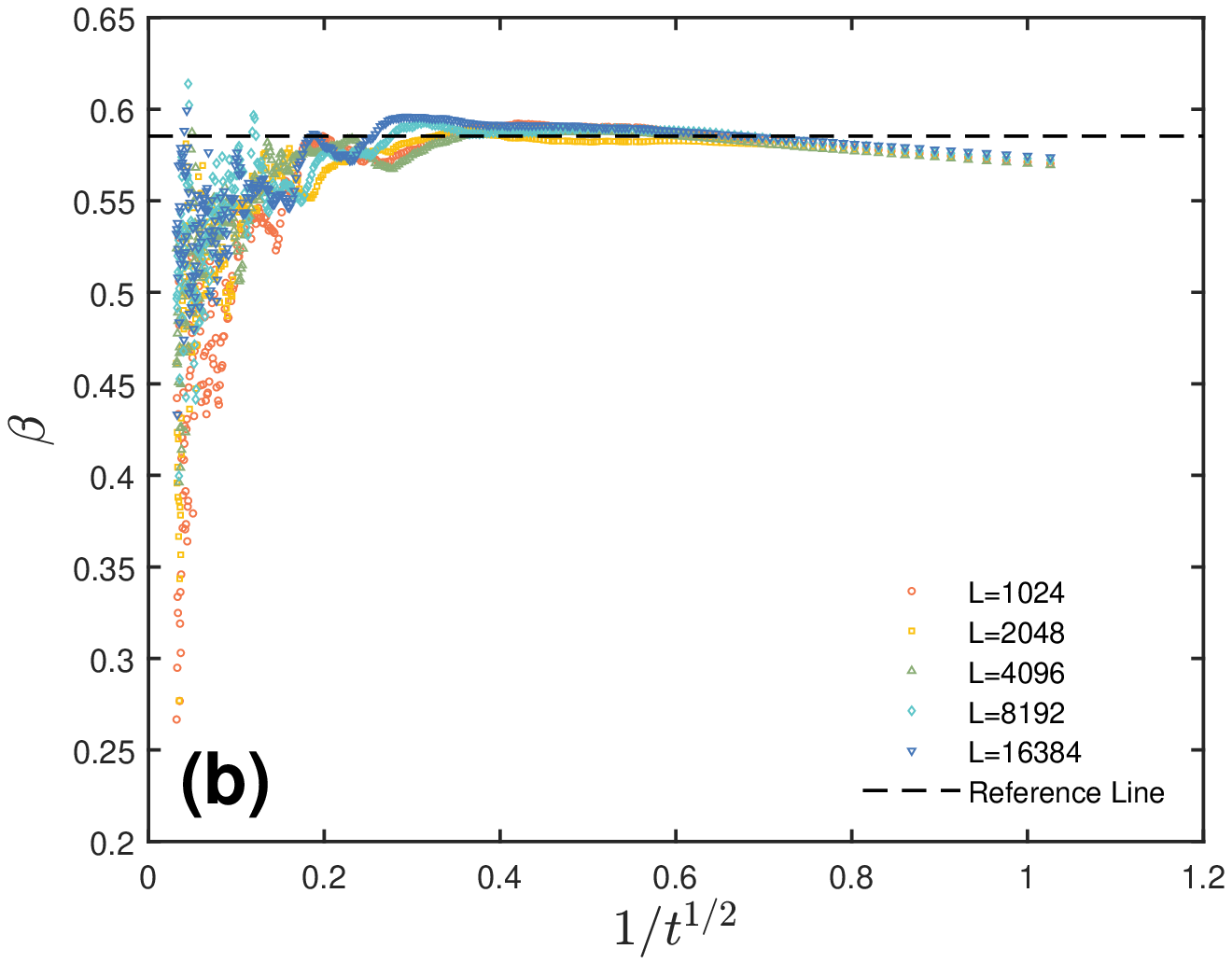}

{Figure A.1: Local slopes versus $1/t^{1/2}$ of the surface growth for different sizes ($L = 1024, 2048, 4096, 8192, 16384$) based on these two numerical schemes: (a) PS and (b) LS. Averaging was done over 100 independent runs.}
\end{figure}

\begin{figure}
\centering
\includegraphics[scale=0.65]{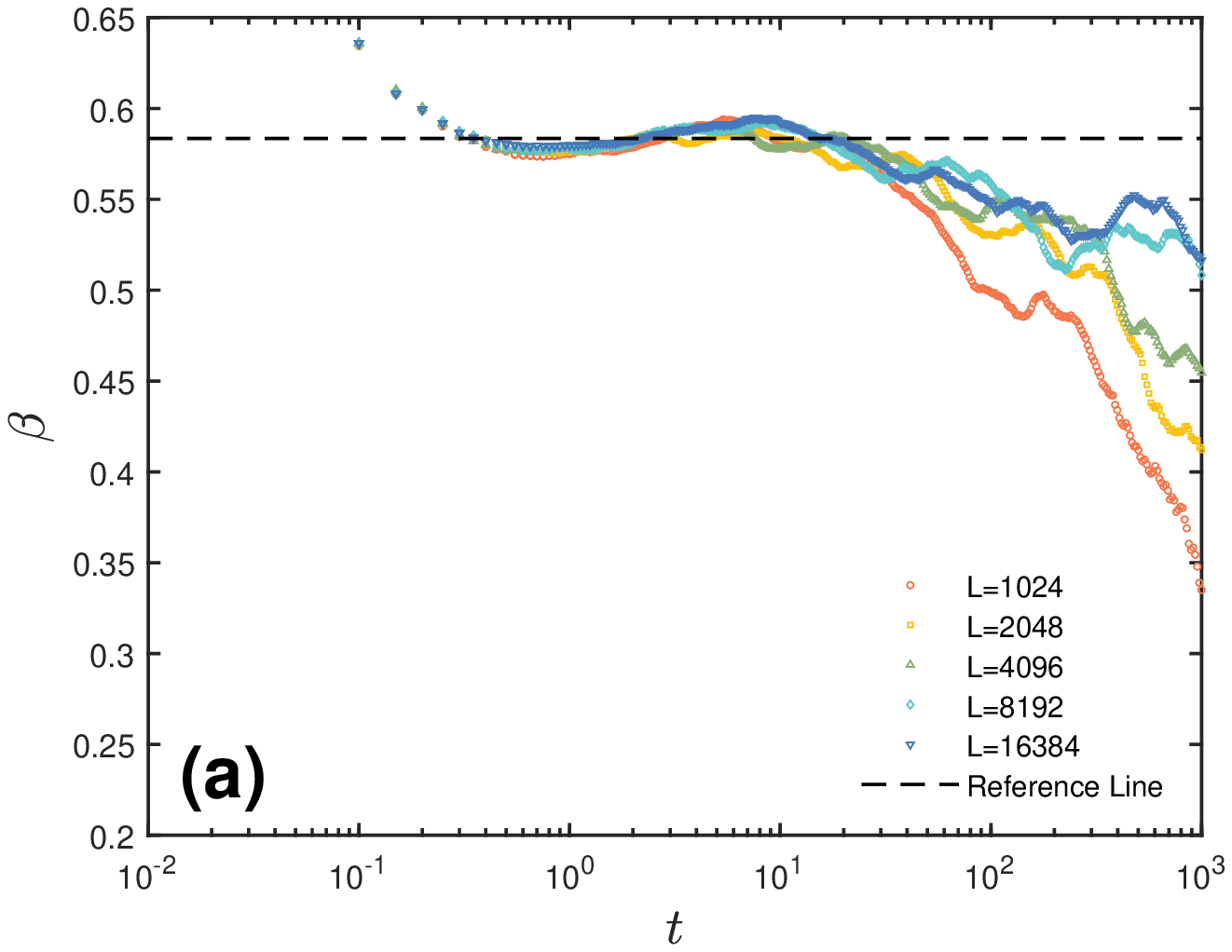}
\includegraphics[scale=0.65]{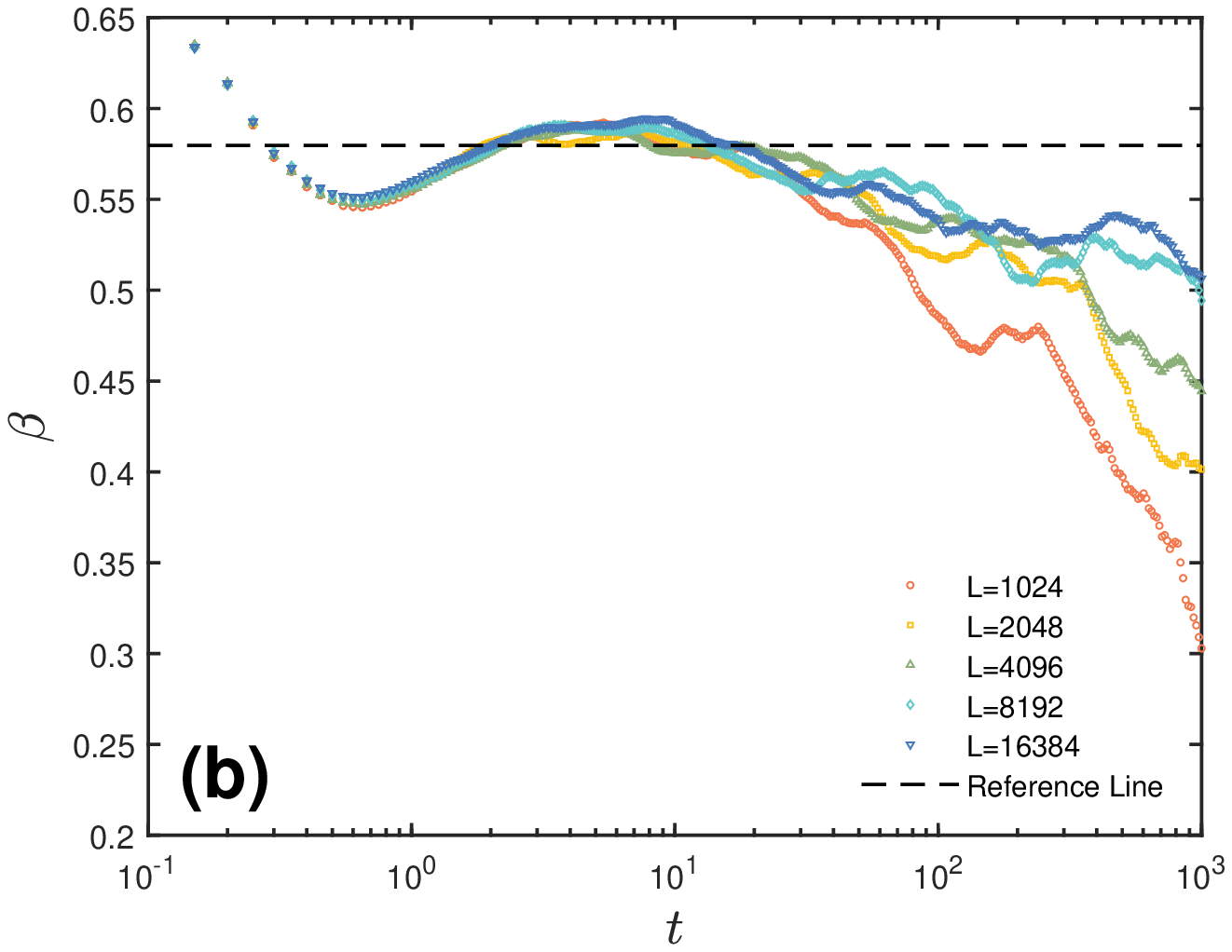}

{Figure A.2: Local slopes versus growth times using (a) PS and (b) LS numerical schemes. All parameters chosen here are the same as those in Fig. A.1.}
\end{figure}

\end{document}